\RequirePackage{snapshot}
\documentclass[runningheads]{llncs} 

\usepackage{graphicx}
\graphicspath{{./}{figures/}}
\usepackage[usenames,dvipsnames,svgnames,table]{xcolor}
\usepackage{mathtools}
\usepackage{amssymb,amsmath}
\usepackage{ifthen}
\usepackage{color}
\usepackage{xspace}
\usepackage{listings}
\usepackage{float}
\usepackage{makecell}
\usepackage{wrapfig}
\usepackage{lipsum}
\usepackage{array}
\usepackage[utf8]{inputenc}
\usepackage{textcomp}
\usepackage{enumitem}
\usepackage{booktabs}
\usepackage{multirow}
\usepackage[final]{pdfpages}
\usepackage{tikz}
\usepackage[moderate]{savetrees}
\usetikzlibrary{decorations.pathmorphing, patterns}
\usepackage[colorlinks=true,linkcolor=blue,urlcolor=blue,citecolor=blue]{hyperref} %
\usepackage{subcaption}
\usepackage{pifont}

\usepackage{wrapfig}
\usepackage[labelfont={small,bf},textfont=small]{caption}
\usepackage[abbreviations]{foreign}  %
\usepackage[binary-units]{siunitx}

\usepackage{pslatex}

\usepackage{hyphenat}
\newcommand{\tz}{\textsc{TrustZone}\xspace}
\newcommand{\arm}{\textsc{Arm}\xspace}
\newcommand{\optee}{\textsc{Op-Tee}\xspace}
\newcommand{\gp}{\textsc{GlobalPlatform}\xspace}

\begin{document}
\mainmatter
\title{Developing Secure Services for IoT with OP-TEE:\\A First Look at Performance and Usability\thanks{This is a pre-print of an article published in ``Distributed Applications and Interoperable Systems'' (DAIS) 2019. The final authenticated version is available online at: \url{https://doi.org/10.1007/978-3-030-22496-7_11}}}
\titlerunning{Developing Secure Services for IoT with OP-TEE} %
\author{Christian Göttel, Pascal Felber, \and Valerio Schiavoni}
\institute{
University of Neuch\^atel, Switzerland,
\email{first.last@unine.ch}
}
 
\maketitle
\begin{abstract}
The implementation, deployment and testing of secure services for Internet of Things devices is nowadays still at an early stage.
Several frameworks have recently emerged to help developers realize such services, abstracting the complexity of the many types of underlying hardware platforms and software libraries.
Assessing the performance and usability of a given framework remains challenging, as they are largely influenced by the application and workload considered, as well as the target hardware.
Since 15 years, \arm processors are providing support for \tz, a set of security instructions that realize a trusted execution environment inside the processor.
\optee is a free-software framework to implement trusted applications and services for \tz.
In this short paper we show how one can leverage \optee for implementing a secure service (\ie, a key-value store). 
We deploy and evaluate the performance of this trusted service on common Raspberry Pi hardware platforms.

We report our experimental results with the data store and also compare it against \optee's built-in secure storage.

\end{abstract}

\section{Introduction}
\label{sec:intro}

Despite the availability of security-oriented instruction sets in consumer-grade processors, high-level frameworks that can help developers use such extensions are still at an early stage.
Moreover, little has been said regarding the performance and usability of these frameworks.
This is unfortunate given that the large majority of devices featuring \arm processors (mobile and not) feature the \tz extensions, introduced since 15 years~\cite{armeverywhere}, and are constantly being improved with new processor revisions.
For instance, \arm recently~\cite{armsvirt} updated its ARMv8.4 architecture of application processors enabling virtualization in the secure world.
The introduction of virtualization in the secure world better improves the isolation of components and resources, and it is expected to boost the trusted applications (TA) ecosystem in developing and using common standards and APIs.

It is only very recently that the first open-source tools aiming to exploit these capabilities have emerged.
Notable examples include Linaro ARM Trusted Firmware~\cite{linaro-trusted-firmware}, ARM GNU Toolchain~\cite{arm-toolchain}, Android's Trusty~\cite{android:trusty}, Trustonic's Kinibi~\cite{trustonic:kinibi}, NVIDIA's TLK~\cite{nvidia:tlk} and finally Linaro's \optee~\cite{optee}.

A major challenge for developers of trusted applications resides in the complexity of the secure platforms themselves.
Despite the existence of standards and APIs, trusted applications remain OS-specific because of the custom libraries provided by the different vendors.
Theses libraries are specialized for the various processors and are required to access secure storage and processing elements.
They rely on drivers shipped with the hardware by the silicon manufacturer.
Furthermore, dispatching trusted OSs requires trusted OS-specific code in the firmware, which adds up to the issue.
This greatly hinders the portability of trusted applications across different trusted OSs and drives TA developers toward implementing and supporting several versions of trusted OS-specific TAs.

In this paper, we focus on a specific framework, \optee~\cite{optee}, which has gained much attraction recently and is arguably the most mature open-source framework for developing trusted application with \arm's \tz extensions.
We describe its architecture and features, and we evaluate its usability and performance by developing a simple key-value store.
We also execute \optee's secure storage benchmark and report our results.
This preliminary study bring insights into the benefits of such framework, able to hide the complexity of the underlying vendor-specific libraries and processor, as well as their performance and overhead.

\begin{figure}[t]
  \centering
    \includegraphics[scale=0.48]{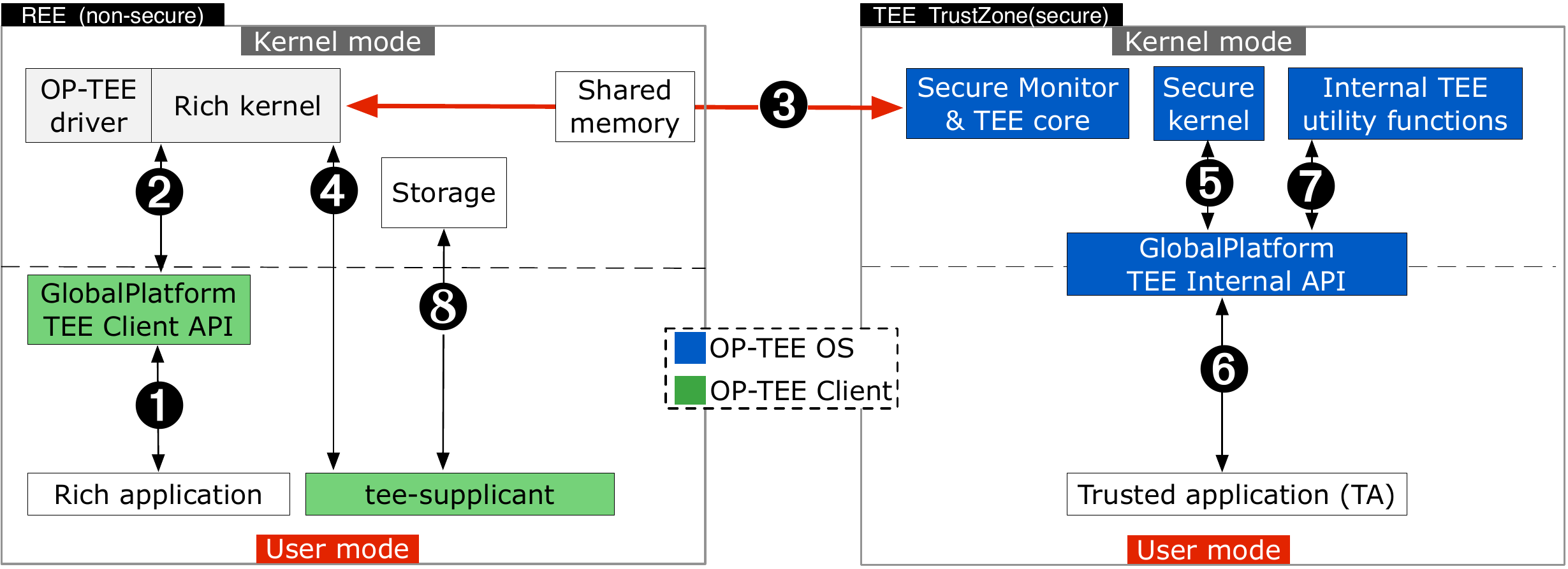}
  \caption{Organization of components within \tz and interaction with \optee}
  \label{fig:tz}
\end{figure}

\section{Background}
\label{sec:background}

\subsection{TrustZone in a Nutshell}
\label{subsec:tz}

The \tz technology is available in \arm processors since 2003~\cite{armsvirt}. 
It is a hardware-enforced mechanism isolating a \emph{secure world} (trusted) from a \emph{normal world} (untrusted), which includes all components within the SoC as well as peripherals. 
Thus, \tz provides secure endpoints and enables device root-of-trust.
Software running in the normal world is unable to directly access secure components and resources. 
When booting up a \tz-enabled SoC, secure firmware is the first software component executed at \emph{exception level 3} (EL3). 
The secure firmware code is responsible for initializing the platform, installing the trusted \emph{operating
  system} (OS) and routing secure monitor calls.
The trusted OS consists of a small and secure kernel to execute \emph{trusted applications} (TA). 
Once the secure world is set up, the normal world OS is booted in parallel to the trusted OS running in the secure world. 
Worlds can be switched via a software-based \emph{secure monitor} (ARMv8-A) or in hardware (ARMv8-M)~\cite{arm:tzv8m}. 
The secure monitor acts as a gateway and runs at the highest privilege level EL3~\cite{armv8}.

\subsection{The GlobalPlatform Specifications for TEEs}
\label{subsec:gp}

The main specifications for secure digital services and devices are published by industry associations~\cite{gp:home,tcg:home}.
In our study, we focus on the \gp specifications for \tz.
A \emph{rich execution environment} (REE) is an execution environment that involves at least one device and all its components or an OS, excluding any trusted or secure component.
In contrast, a \emph{trusted  execution environment} (TEE) provides a level of security to protect against attacks and secures data access. 
The TEE executes alongside the REE, but is shielded from it.
A trusted application executes inside a TEE and exposes secure services to applications in the REE. %
\emph{Trusted storage} is a hardware or cryptographically-protected device capable of storing data~\cite{gp:sysarch}. 
Data can be exchanged between an application in the REE and a TA by three types of shared memory:
\emph{whole} (an entire memory region and is allocated by the TEE),
\emph{partial} (only a a subset of the \emph{whole} with a specified offset), and
\emph{temporarily}, for which a memory buffer region allocate by the application in the REE temporarily shared with the TA for the duration of the API call~\cite{gp:client}.

\begin{table}[t]
  \centering
  \setlength{\aboverulesep}{0pt}
  \setlength{\belowrulesep}{0pt}

  \rowcolors{1}{gray!10}{gray!0}
  \begin{tabular}{>{\kern-\tabcolsep}lll<{\kern-\tabcolsep}}
    \toprule\rowcolor{gray!25}
    \multicolumn{1}{c}{\textbf{Device}} & \multicolumn{1}{c}{\textbf{QEMU}} & \multicolumn{1}{c}{\textbf{Raspberry}} \\
    \midrule%
    CPU & Intel Xeon E3-1270 v6 & Broadcom BCM2837 \\
    \rowcolor{gray!10}
    CPU Frequency & \SI{3.8}{\GHz} & \SI{1.2}{GHz} \\
    \rowcolor{gray!0}
    Memory & \SI{63}{\gibi\byte} DDR4 &
    \SI{944}{\mebi\byte} LPDDR2 \\
    \rowcolor{gray!10}
    Memory Frequency & \SI{2400}{\MHz} & \SI{900}{\MHz} \\
    \rowcolor{gray!0}
     & Samsung &
    Transcend micro SDHC \\
    \rowcolor{gray!0}
    \multirow{-2}{*}{Disk} & MZ7KM480HMHQ0D3 & UHI-I Premium \\
    \rowcolor{gray!10}
    Disk Size & \SI{480}{\giga\byte} & \SI{16}{\giga\byte} \\
    \rowcolor{gray!0}
    Disk Read Speed & \SI[per-mode=symbol]{528.33}{\mega\byte\per\second} &
    \SI[per-mode=symbol]{90}{\mega\byte\per\second} \\
    \bottomrule
  \end{tabular}
  \caption{Comparison of platforms}
  \label{tab:platform}
\end{table}

\subsection{The OP-TEE Framework}
\label{subsec:optee}

\optee~\cite{optee} is a TEE implementation of \gp specifications on top of \tz.
It can be used alongside a Linux-based distribution running in the REE.
TAs are single-threaded executables stored inside the REE.
Users develop TAs without having to recompile the entire framework.
\optee does not provide mechanisms to verify the integrity of a TA, and consequently it exposes TAs to the untrusted REE. 
Upon modification, this can compromise the integrity or protection of the TEE. 
Alternatively, TAs can be directly integrated into \optee as \emph{pseudo TAs}.
Pseudo TAs run inside the \optee OS as secure privileged-level services without access to \gp's Internal Core API.
Thus, pseudo TAs can only use \optee's core Internal API. %

Secure storage allows applications to offload data from a TA to either the REE file system or a \emph{replay protected memory block} (RPMB) partition of an \emph{embedded multi-media controller} (eMMC) device using the Internal Core
API.
By default, the \optee OS is configured to use the RPMB~\cite{optee:secstor} if available. 
The secure storage is accessible and visible only to the TA that created it.

\section{Usability}
\label{sec:usability}

The communication between an application in the normal world and a TA evolves around functions handling the context, session, command and shared memory as shown in~\autoref{fig:tz}.
This facilitates interoperability between different \gp API compatible TEE implementations and allows REE applications to set up multiple contexts.
A context is initialized by referencing the device file~\ding{182} connecting to the TEE driver~\ding{183}.
TAs are identified by a \emph{universally unique identifier} (UUID), which is referred to when setting up a session to a TA~\ding{184}.
To set up a session, \optee will load the TA from the normal world to the secure world with the help of \texttt{tee-supplicant}~\ding{185}.
The \texttt{tee-supplicant} is a daemon running in the normal world used by \optee to request services from the REE.
These steps are skipped when a session to a pseudo TA is established.
A TA can initialize and set up its environment upon TA creation and session establishment (\ding{186} \&~\ding{187}).
From this point on, the REE application can request services from the TA by invoking commands.
These commands can pass up to four parameters, which are either values or references to shared memory regions.
Values are pairs of unsigned \SI{32}{\bit} integers.
Shared memory regions are allocated, registered and released through \gp API calls in \texttt{libteec}.
Without the availability of \texttt{libteec}, developers would have to communicate directly with the kernel driver through \texttt{ioctl} system calls.

In \optee, TAs can use services accessible through \gp Internal Core API~\ding{187} implemented in \texttt{libutee}.
TAs are statically linked against \texttt{libutee}, which wrapps the API functions around assembler macros to \optee OS system calls.
The library provides interfaces to secure storage~\ding{189}, time, arithmetic and cryptographic operations~\ding{188}.
The secure storage API encrypts data objects by the use of a secure storage service.
The encryption process involves three keys: \emph{secure storage key} (SSK), \emph{trusted application storage key} (TSK) and \emph{file encryption key}.
The SSK is generated from the \emph{hardware unique key} and is used to derive TSKs.
Each TA has a TSK that is generated from the SSK and the TA's UUID.
Both SSK and TSK are generated using HMAC SHA256 algorithm~\cite{optee:secstor}.
Finally, every created file a FEK is generated from the pseudo random number generator.
The encrypted data objects are then transferred to the \texttt{tee-supplicant} by a series of remote procedure calls and stored in a special file.
\optee further provides TAs with libraries for TLS and SSL protocols (\texttt{libmbedtls}~\cite{mbedtls}), arithmetic (\texttt{libmpa}) and a subset of ISO C functions (\texttt{libutils}).
Once the REE application has no further service requests, the session is terminated and the context is destroyed.

\begin{figure}[t]
  \centering
  \begin{subfigure}[b]{0.48\textwidth}
    \centering
    \includegraphics[trim={5px 30px 35px 205px},clip,width=\textwidth]{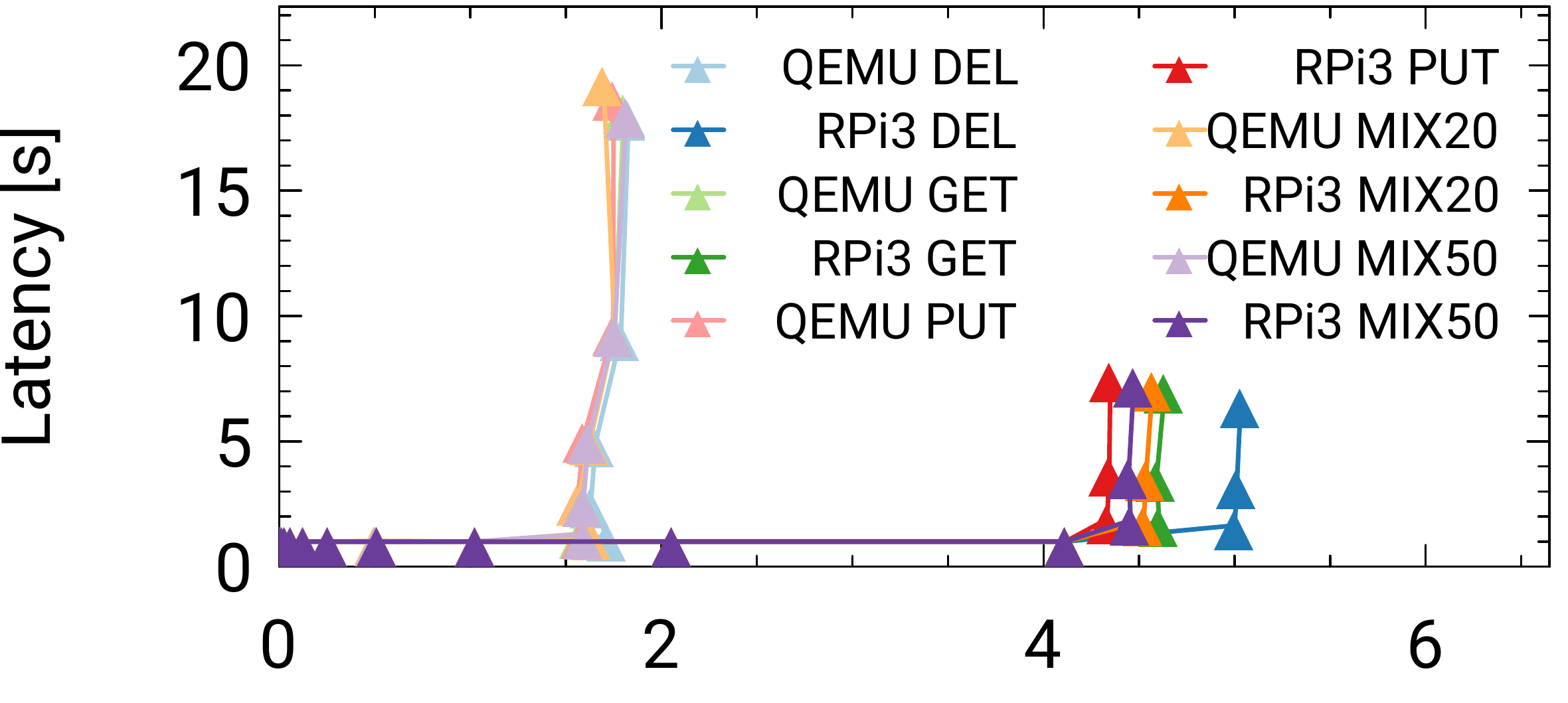}
    \caption{Partially SHM}
  \end{subfigure}%
  \begin{subfigure}[b]{0.48\textwidth}
    \centering
    \includegraphics[trim={100px 30px 0 205px},clip,width=0.915\textwidth]{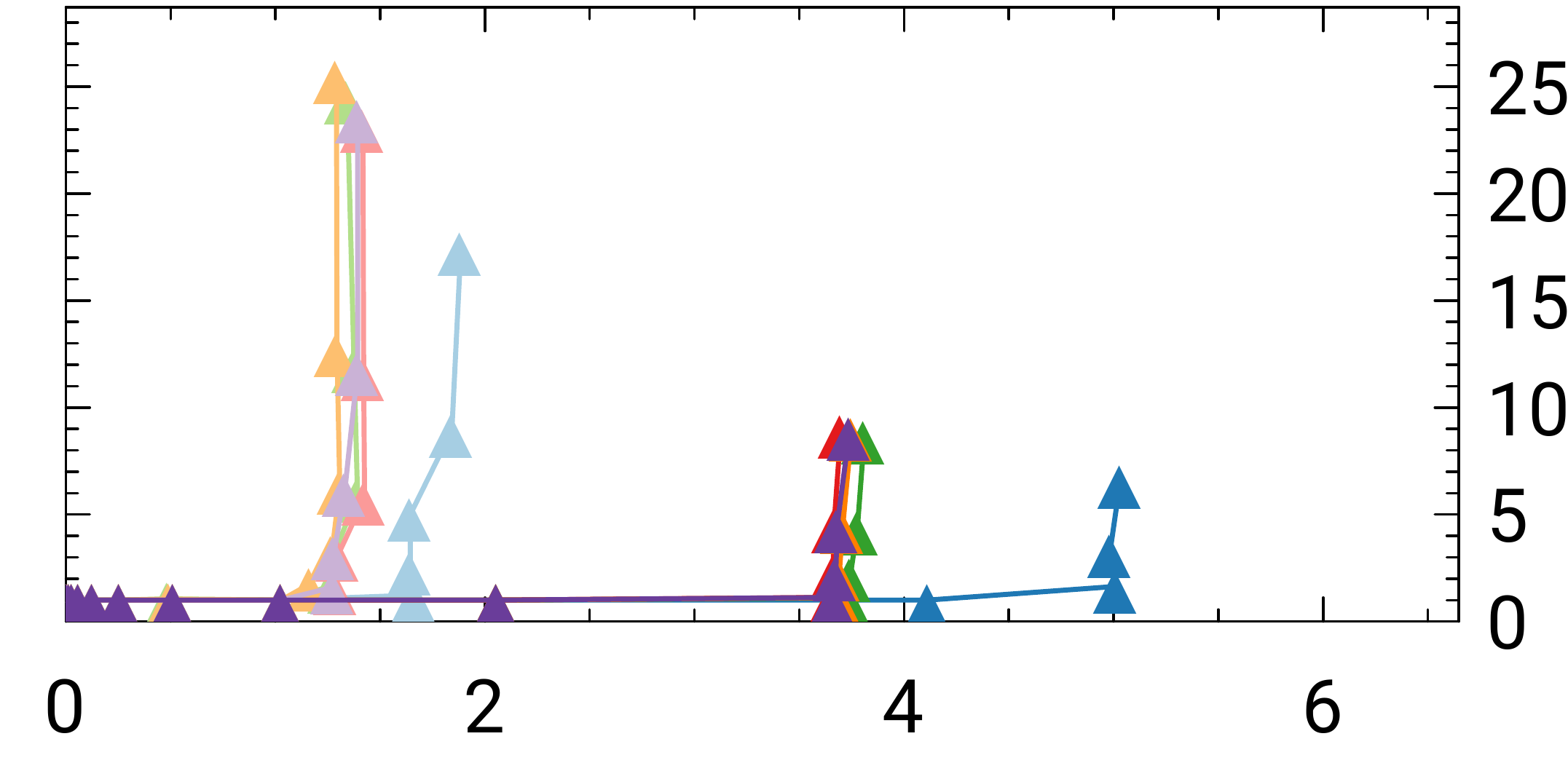}
    \caption{Temporarily SHM}
  \end{subfigure}\\\vskip 1em%
  \begin{subfigure}[b]{0.48\textwidth}
    \centering
    \includegraphics[trim={5px 0 35px 205px},clip,width=\textwidth]{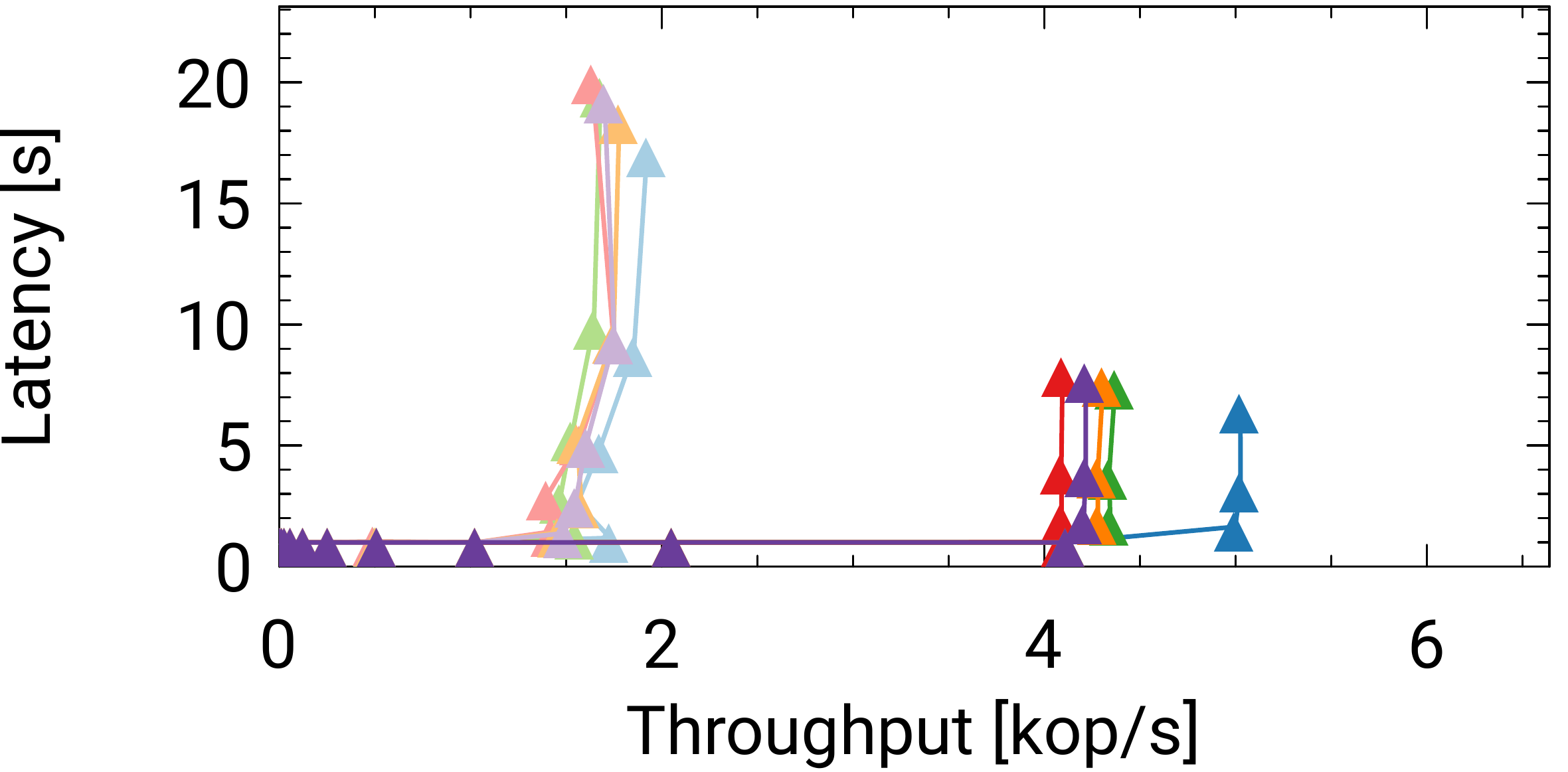}
    \caption{Whole SHM}
  \end{subfigure}%
  \begin{subfigure}[b]{0.48\textwidth}
    \centering
    \includegraphics[trim={100px 0 0 205px},clip,width=0.915\textwidth]{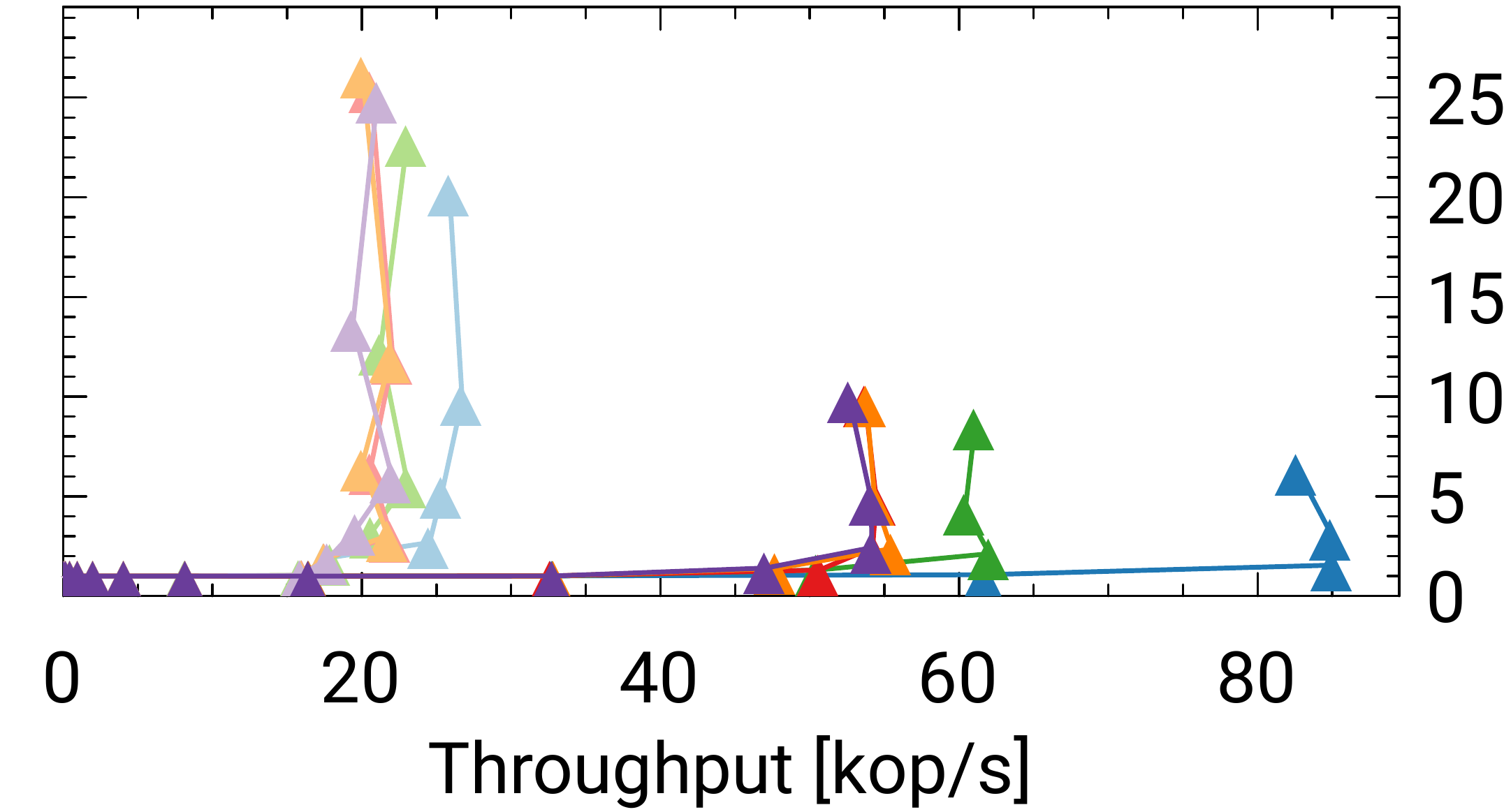}
    \caption{Inside REE}
  \end{subfigure}
  \caption{Throughput-latency plots of shared memory types for key-value
    TA in TEE and REE.}
  \label{fig:kv}
\end{figure}

\section{Performance Evaluation}
\label{sec:evaluation}

\subsection{Setup}
The key-value store and \optee's Sanity Testsuite v3.2.0~\cite{optee:xtest} were deployed on the two platforms: Dell PowerEdge R330 Server and Raspberry Pi 3B v1.2.
The Dell PowerEdge R330 is running Ubuntu 18.04.1 LTS with the 4.15.0-43-generic Linux kernel and is used to emulate the Raspberry Pi 3B platform with QEMU v2.12.0~\cite{qemu}.
A comparison of the two platforms can be found in~\autoref{tab:platform}.
\optee provides a build environment which, by default, deploys and emulates its OS on an ARM Virtual Machine \texttt{virt} using a Cortex-A57 with no more than two cores.
The deployment was changed to match the specification of the Raspberry Pi 3B platform as close as possible.

\subsection{Shared Memory}

We have ported a simple key-value store to a TA, in order to evaluate the overhead and performance of different types of shared memory.
As basis, we used a modified version of the hash table implementation of \texttt{kazlib} v1.20~\cite{kazlib}, removing support for contexts and dynamic tables.
The hash table is static, uses separate chaining to resolve collisions, applies a modular hashing and has \num{251} chains.
We time every \texttt{DEL} (delete), \texttt{GET} and \texttt{PUT} operation for each benchmark by referring to \texttt{CLOCK\_MONOTONIC} in the REE.
Operations are uniformly distributed and issued \num{256} times at a rate of \numrange{1}{32768} operations per second.

When using whole or partially shared memory, the REE application requests a shared memory region of \SI{512}{\kibi\byte} and fills it with random data from \texttt{/dev/urandom}.
Similarly, the REE application allocates and initializes a \SI{512}{\kibi\byte} buffer used as temporarily shared memory. Before every invocation of a key-value operation, a random offset into the shared memory region is computed, which is also used as key.
A chunk size of \SI{1}{\kibi\byte} beginning at the random offset is used as data object.
The \texttt{PUT} benchmark starts with an empty hash table.
The \texttt{DEL} and \texttt{GET}  benchmarks start with a pre-populated hash table of \num{256} data objects.
Finally, the mixed benchmark (ratio of \texttt{GET} and \texttt{PUT} operations) begins with a pre-populated hash table relative to the percentage of \texttt{GET} operations.

\autoref{fig:kv} shows throughput and latency for the different shared memory types and for running the key-value store entirely in the REE.
On the QEMU platform, the operations do not separate as well as on the Raspberry platform; we assume due to reaching an I/O bound.
The operations on the Raspberry platform separate as expected according to their throughput (lowest to highest): \texttt{PUT}, \texttt{MIX50}, \texttt{MIX20}, \texttt{GET}, and \texttt{DEL}.
The overhead of the \texttt{PUT} operation is due to memory allocation, memory copy and object insertion.
The \texttt{GET} operation looks up a data object and copies it to shared memory, resulting in a lower overhead.
The higher the portion of \texttt{PUT} operations in the \texttt{MIX} benchmarks is, the slower the average operation speed becomes.
Thus, \texttt{MIX50} (\SI{50}{\percent} \texttt{PUT} operations) has a lower average throughput than \texttt{MIX20}.
The \texttt{DEL} operation looks up a data object and frees its memory, avoiding time consuming memory operations.
Comparing TEE throughput against REE throughput yields a \SIrange[range-units=single]{12}{14}{\times} overhead on the QEMU
platform and a \SIrange[range-units=single]{12}{17}{\times} overhead on the Raspberry platform.
A similar experiment was conducted in~\cite{scitepress}, where they compared the time spend in normal and secure world when invoking a noop operation.

\subsection{Secure Storage}
The secure storage benchmark is part of the \optee sanity test suite adhering to the \emph{Trusted Storage API for Data and Keys} described in~\cite{gp:core}.
Neither of the platforms is equipped with an eMMC, for which reason the secure storage has to be offloaded to the REE file system.
The benchmark executes three commands \texttt{WRITE}, \texttt{READ}, and \texttt{REWRITE}, for data sizes in the range of \SI{256}{\byte} to \SI{1}{\mebi\byte}, that are accessed in chunks of at most \SI{1}{\kibi\byte}. 
The \texttt{REWRITE} command first reads data from an object, resets the cursor and writes the data back to the same object.
The data to be stored in the secure storage is allocated and filled with scrambled data within the TEE.

\begin{figure}[t]
  \centering
  \includegraphics[width=\textwidth,trim={0 0 0 210px},clip]{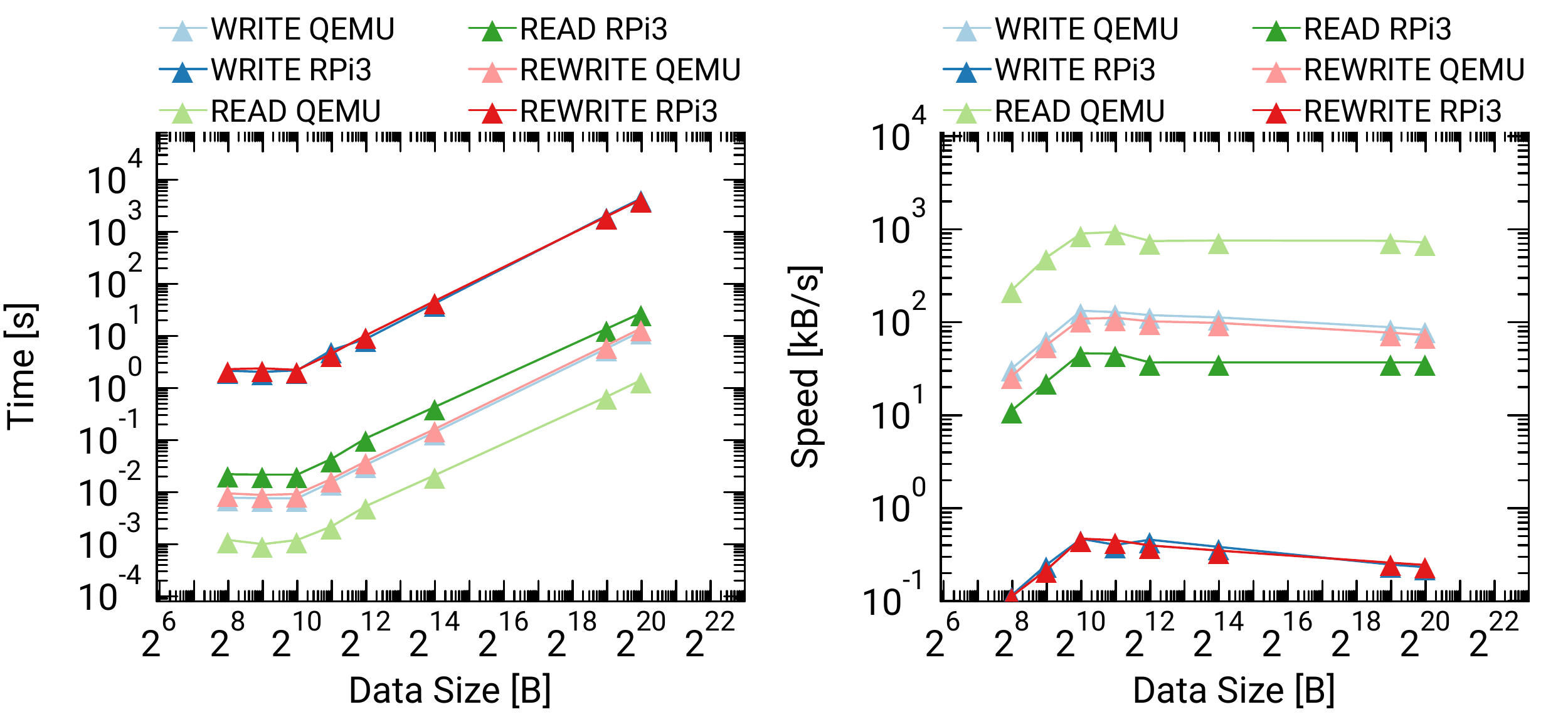}
  \caption{Secure storage benchmark execution time and throughput}
  \label{fig:ssspeed}
\end{figure}

\autoref{fig:ssspeed} shows the overhead of accessing data in chunks of \SI{1}{\kibi\byte} in the secure storage.
In general, the overhead becomes more significant with increasing data sizes, more precisely once the data size exceeds the chunk size.
Maximum speed is achieved when the data size equals the chunk size.
Overall, the \texttt{REWRITE} command has the highest overhead, because it basically executes the \texttt{READ} and \texttt{WRITE} commands in one batch.

\section{Concluding Remarks}
\label{sec:conclusion}

Development of secure services benefits from well established APIs and standards.
\optee has implemented several of \gp's specifications and APIs and provides common interfaces for secure services.
We have ported a simple key-value store to a TA and we have studied the performance and usability of secure storage and shared memory.
The results of our benchmarks have shown that requesting services from TAs in \tz on ARMv8-A using \optee incurs a significant overhead compared to service execution in the normal world.
Limiting the space available to a TA is sensible, in order to minimize the trusted computing base.
However, the default memory limit of \SI{1}{\mebi\byte} for TAs in \optee becomes a major  inconvenience with respect to secure storage and shared memory. 

Generating the SSK in \optee requires the HUK.
However, most platforms lack of documentation to access or obtain the HUK.
\optee avoids this issue by considering a static string value instead of the HUK.
This alternative can potentially weaken the cryptographic protection of the objects stored in the REE file system of the secure storage.
TEEs would greatly benefit from unrestricted access to HUKs and could so improve the protection of trusted storage.

We expect the trusted application ecosystem to improve portability of TAs among TEEs.
Furthermore, we hope that our evaluation of usability and performance of TAs provides
deeper insight into future development of trusted services.

\section*{Acknowledgments}
The research leading to these results has received funding from the European Union's Horizon 2020 research and innovation programme under the LEGaTO Project (\href{https://legato-project.eu/}{legato-project.eu}), grant agreement No~780681.

{\footnotesize
  \bibliographystyle{splncs04}
  \bibliography{biblio}
}

\end{document}